\documentclass{kluwer}    
\usepackage[dvips]{graphicx}

\newdisplay{guess}{Conjecture}

\newcommand{\gs}{{_>\atop^{\sim}}}

\newcommand{\ltsima}{$\; \buildrel < \over \sim \;$}
\newcommand{\simlt}{\lower.5ex\hbox{\ltsima}}                                  

\newcommand{\cgs}{${\rm erg~cm}^{-2}\ {\rm s}^{-1}$}

\def\heao1{{\it HEAO-1\/}}

\def\xmm{{XMM-{\it Newton\/}}}

\def\apj{{\it ApJ}}
\def\apjs{{\it ApJS}}
\def\aap{{\it A\&A}}
\def\aj{{\it AJ}}

\def\etal{{\it et al.}}

\begin{document}

\begin{article}
\begin{opening}         
\title{The density and masses of obscured Black Holes}
\author{Andrea \surname{Comastri}}  
\runningauthor{Comastri A. \& Fiore F.}
\runningtitle{Obscured AGN}
\institute{INAF--Osservatorio Astronomico di Bologna} 
\author{Fabrizio \surname{Fiore}}  
\institute{INAF--Osservatorio Astronomico di Roma} 
\date{March 15, 2004}

\begin{abstract}
Recent {\it Chandra} and XMM--{\it Newton} surveys 
have uncovered a large fraction of the obscured AGN responsible 
of the hard X--ray background. One of the most intriguing results of 
extensive programs of follow--up observations concerns 
the optical and near--infrared properties of the 
hard X--ray sources counterparts. More specifically, for a significant 
fraction of hard X--ray obscured sources the 
AGN responsible of the high X--ray luminosity remains elusive 
over a wide range of wavelengths from soft X--rays to near--infrared.
This very observational result opens the possibility to investigate the
host of bright obscured quasars in some detail.
Here we briefly report on some preliminar results obtained 
for a small sample of elusive AGN in the {\tt HELLAS2XMM} survey.

\end{abstract}

\end{opening}           

\section{Introduction} 

Hard X-ray surveys represent an efficient probe to unveil the
super-massive black hole ({\tt SMBH}) accretion activity, which is
recorded in the cosmic X-ray background ({\tt XRB}) spectral intensity.  
The advent of hard X-ray (i.e., 2--10~keV) imaging instruments, from {\tt
ASCA} to {\it BeppoSAX} and, more recently, {\it Chandra} and \xmm,
has provided a dramatic advance in the field of X--ray surveys.

The combination of deep/ultra-deep X--ray surveys with {\it Chandra}
(Alexander et al. 2003; Giacconi et al. 2002) and the shallower
surveys with \xmm\  (Baldi et al. 2002) has allowed to resolve
\hbox{$\approx$~80\%} of the {\tt XRB} in the \hbox{2--10~keV} band and to
unveil classes of cosmic sources which were previously unknown or
marginally represented by a few ambiguous and sparse cases.

Within this context, the {\tt HELLAS2XMM} survey plays an important 
role. Using suitable \xmm\ archival observations,  
this project aims at covering 4 square degrees of sky down to X--ray fluxes
of the order of $10^{-14}$ \cgs, sampling the bright 
tail of the X--ray luminosity function.  
The key scientific issue is a solid 
estimate of the luminosity function and evolution of the 
obscured AGN responsible for a large fraction of the {\tt XRB}.
In order to fulfill the goals of such an ambitious objective,  
the spectroscopic identification of a large number of hard X--ray selected
sources is mandatory. Unfortunately the identification of 
a sizable fraction of optically faint obscured AGN is already challenging
the capabilities of the 8--10 m class telescopes calling for alternative
approaches based on multiband optical photometry, 
detection of redshifted iron K$\alpha$ lines, or statistical 
methods. Some of the results obtained by our group in these 
regards,  based on the multiwavelength observations of 
the {\tt HELLAS2XMM} 1 degree field, will be briefly reported.

\section{Optically obscured AGN}

\begin{figure} 
\includegraphics[width=0.9\textwidth]{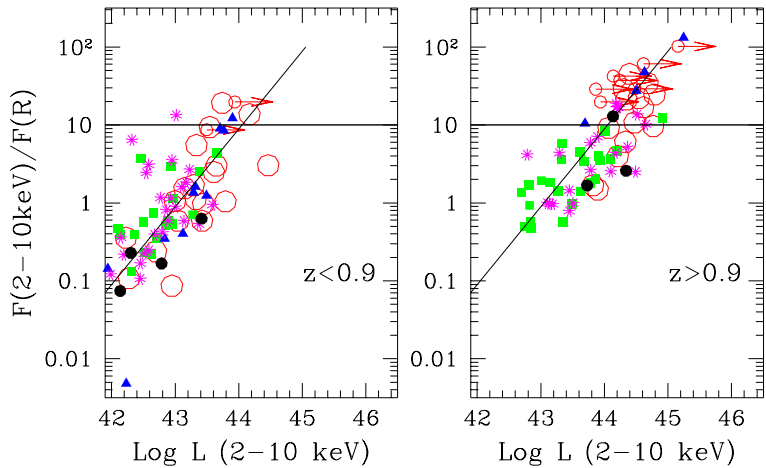}
\caption[]{X/O versus the X-ray luminosity optically obscured AGN (without
broad lines) in two redshift bins.  HELLAS2XMM = open circles
(arrows mark photo-z by Mignoli et al. 2004); CDFN = filled squares;
CDFS = stars; LH = filled triangles; SSA13 = filled circles, ).  Solid
lines mark loci of constant R band magnitude.  The orizontal lines
mark the level of X/O=10, $\sim20\%$ of the sources in the combined
sample have X/O higher than this value. The diagonal line in the right
panel is the best log(X/O)--log(L$_{2-10keV}$) linear regression in
Fiore et al. (2003).}
\label{xolxz}
\end{figure}

About 20\% of the sources detected in recent hard X-ray surveys have a
large X-ray \hbox{(2--10~keV)} to optical ($R$-band) flux ratio
(X/O $>10$), i.e., ten times or more higher than typically observed in
optically selected AGNs (Fiore et al. 2003; Alexander et al. 2001) The
fraction of these sources seems to remain constant over $\approx$~3
decades of X-ray flux. Fiore et al. (2003) discovered a striking
correlation between X/O and the 2--10 keV luminosity for the sources
having strongly obscured nuclei in the optical band, i.e. not
showing broad emission lines. The optical R band light of these
objects is dominated by the host galaxy light, which spans a
luminosity range much smaller (about one fourth) than the X-ray
luminosity, giving rise to the above correlation.
Figure \ref{xolxz} shows X/O as a function of the 2--10 keV luminosity
for a sample of optically obscured AGN in two redshift bins, below and
above $z$=0.9. The correlation between X/O and the 2--10 keV luminosity,
though characterized by a not negligible scatter,  
is very strong in both redshift bins. 
At redshifts higher than  0.9 most of the point with 
L(2--10 keV) $>$ 10$^{44}$ erg s$^{-1}$, corresponding to high
luminosity, highly obscured type 2 QSO, are from the {\tt HELLAS2XMM}
sample (Figure~1 right panel). 
The reason is that at the flux limit of the {\tt HELLAS2XMM} sample
several sources with X/O$\gs10$ have optical magnitudes up to R $\sim$ 
24--25,
bright enough to obtain reliable spectroscopic redshifts, while in the
much deeper but smaller area surveys of the CDFN, CDFS and Lockman
Hole, sources with X/O$\gs10$ have fainter (up to R $\sim$ 29) optical 
magnitudes,
unaccessible to spectroscopy even with 10m class telescopes. Indeed, we
were able to obtain spectroscopic redshifts and classification of 13
out of the 28 {\tt HELLAS2XMM} 1dF sources with X/O$>10$; {\em 8 of them are
type 2 QSO at $z$=0.7--1.8}, to be compared with the similar number
obtained from the combination of the CDFN and CDFS, at the expenses of
a huge investment of VLT and Keck observing time. For other 8
{\tt HELLAS2XMM} sources with X/O$\gs10$ a lower limit on their
redshifts was obtained from
their observed R--K color (Mignoli et al. 2004) and they nicely agree
with the X/O--logL(2--10 keV) correlation (see the small open circles
with arrows in figure \ref{xolxz}).  We conclude that the majority of
the sources with X/O$\gs10$ are type 2 QSO at z$\gs1$, postulated in
the simplest versions of XRB models based on AGN unification schemes
(Comastri et al. 1995).  
Although only 20\% of the X-ray sources have such high
X/O, they may carry the largest fraction of accretion power from that
shell of Universe.  Large area (of the order of several square degrees), 
medium--deep surveys like 
{\tt HELLAS2XMM} may be the most efficent tool to fully uncover this
still rather poorly known component of the AGN population.

\section{X-ray obscured AGN}

\begin{figure} 
\includegraphics[width=0.9\textwidth]{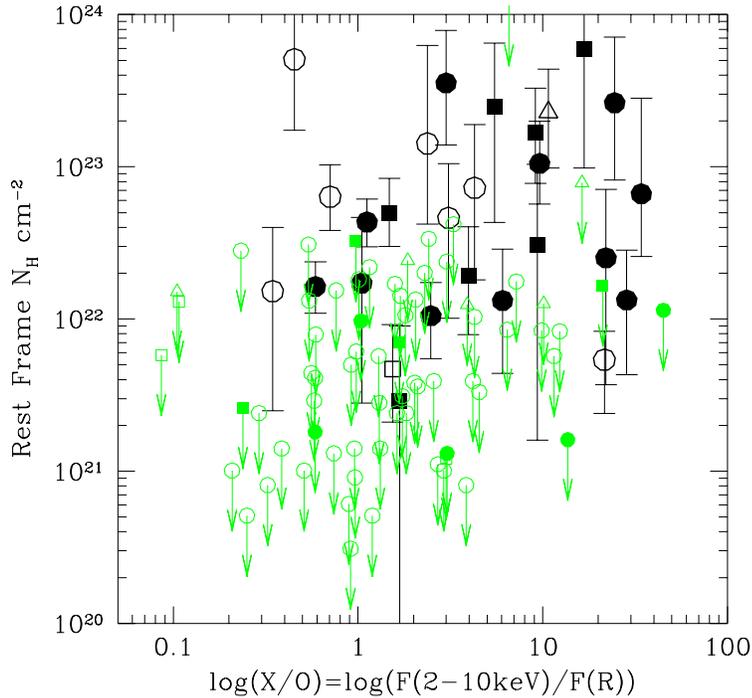}
\caption[]{The best fit rest--frame absorbing columns from Perola et
al. 2004 as a funtion of X/O for the HELLAS2XMM 1df sample.  Different
symbos identify different classification of the optical spectra: open
circles = type 1 AGN; filled circles = type 2 AGN; filled squares =
emission line galaxies; open squares = early--type galaxies. Open
trianges refer to sources without optical spectroscopic
identification. For these sources z=1 was assumed in the X-ray
spectral fitting.}
\label{xonhf}
\end{figure}

The X-ray absorption
properties of the {\tt HELLAS2XMM} 1df sample  have been studied in detail 
by Perola et al. (2004). Figure \ref{xonhf}
shows X/O as a function of the best fit, rest frame absorbing column
for these sources. A strong correlation is clearly present between X/O
and $N_H$, i.e. optically obscured sources tend to be, on average,
obscured also in X--rays, in agreement with Unification Schemes,
although exceptions to this trend do exist. For istance X--ray 
absorption with column densities $N_H \sim$ 10$^{22-23.5}$ in 
broad line type 1 AGN has been reported (see Brusa et al (2003) 
and Perola et al (2004) for more details).

\begin{figure} 
\includegraphics[width=0.9\textwidth]{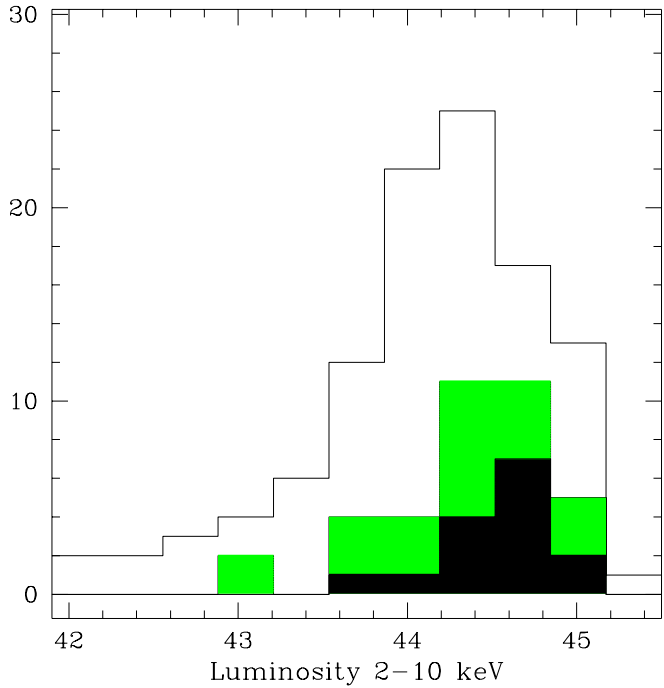}
\caption[]{The distribution of the HELLAS2XMM 1df sources
as a funtion of the 2-10 keV luminosity, for three log$N_H$ bins: 
white histogram =  $<22$, gray histogram  = $22-23$ and black histogram
= $>23$.}
\label{lxhisto}
\end{figure}

The distribution of the {\tt HELLAS2XMM} 1df
sources as a function of the 2--10 keV luminosity is shown in 
Figure \ref{lxhisto} for three logarithimic bins of the absorbing 
column density:
$N_H<22$, $22<N_H<23$ and $N_H>23$. The fraction of obscured sources
in this sample is roughly constant with the luminosity. This is likely to
be due to
the rather narrow flux range (about one decade) covered by the
{\tt HELLAS2XMM} sample. Going to fainter fluxes both lower
luminosities and higher redshifts are sampled. As a consequence,  
the correlations between the
fraction of obscured sources with the luminosity and/or the redshift
start to emerge (Ueda et al. 2003, Hasinger 2003, La Franca et
al. in preparation).

Figure \ref{lxhisto} also shows that the fraction of high luminosity,
highly obscured X-ray obscured objects can be as high as 40\% (see
Perola et al.  2004 for details), which means $\sim48$ deg$^{-2}$ at
the flux limit of the {\tt HELLAS2XMM} survey.

\section{The masses of obscured black holes}

To study the sources with high or extreme X/O in the
{\tt HELLAS2XMM} sample, and faint (R$>24$) optical counterparts in 
more detail, we have
started a pilot program of deep near-infrared photometry
($K^{\prime}<21.5$) using {\tt ISAAC@VLT}.  Ten of the 11 sources with
X/O $>$ 10 observed in the $K^{\prime}$ band were detected ($\langle
K^{\prime}\rangle\approx$~18.4; Mignoli et al. 2004).  Their
optical-to-near-infrared colors are significantly redder than those of
the field galaxy population, all of them being Extremely Red Objects
(EROs; $R-K>5$).

The results of the analysis of the image profiles indicate that a
pointlike nuclear source is present in a few objects only, the
majority of the sources (7 out of 10) being well represented by a de
Vaucoulers profile. A redshift estimate has been obtained from the
measured optical and near--infrared magnitudes (see 
Mignoli et al. 2004 for a detailed description).
Briefly, the observed distribution of R--K colors
of our sources is compared with that of the sub--sample of 
spectroscopically identified 
early--type galaxies in the K20 redshift survey (Cimatti et al. 2002).
A simple stellar population model which provides the reddest colors
at each redshift has been then employed to assign a ``minimum'' 
redshift to each source. Not surprisingly, given their EROs color and
K--band morphology, the 7 objects lie in the range 0.9 $< z <$ 1.5.

\begin{figure} 
\includegraphics[width=0.9\textwidth]{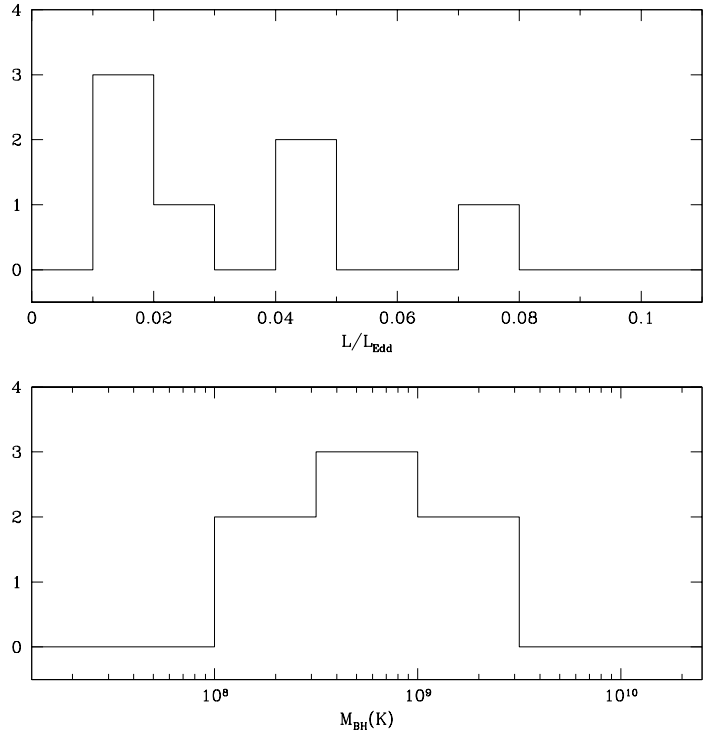}
\caption[]{The Eddington ratio of the 7 AGN for which
the K--band light is dominated by a bulge component (top panel). 
The estimated black hole masses in units of $M_{\odot}$ (lower panel)} 
\label{penG}
\end{figure}

The masses of the obscured {\tt SMBH} powering the X--ray source
and hidden by the host galaxy starlight 
can be estimated from the relation between the black hole mass 
and the K--band luminosity of the bulge component 
as discussed by Marconi \& Hunt (2003).
This approach returns an  upper limit
to the SMBH mass, since it has been assumed that the measured 
K band light is completely dominated by the bulge component. 
The presence of a disk or a residual point--source contribution
escaping detection in the azimuthally--averaged 
brightness profile would lower the bulge luminosity and 
in turn the mass determination.

The mass upper limits coupled with the unabsorbed 2--10 keV luminosities
derived by Perola et al. (2004) and with a prescription 
for the bolometric correction ($k_{bol}$) allow a straightforward 
calculation of the lower limits on the Eddington ratio ($L_{bol}/L_{edd}$).
The correction factor needed to estimate the bolometric luminosity 
from the 2--10 keV band is subject to significant uncertainties 
and a robust estimate ($k_{bol} \sim$ 30) is available only for 
bright unobscured quasars (Elvis et al. 1994). 
Lower values $k_{bol} \sim$ 10--20 are more appropriate 
for lower luminosities Seyfert--like objects (Fabian 2003) and
also appears to better reproduce the  
observed Spectral Energy Distribution of a few 
heavily obscured sources (Comastri 2004). 
Given that we are dealing with lower limits of the Eddington ratio
a conservative value of $k_{bol}$=10 has been assumed. 
The distribution of the SMBH masses and L/L$_{Edd}$ for the 7 galaxies 
are shown in Figure \ref{penG}.
The present estimates obtained through a chain of 
assumptions and neglecting the uncertainties associated to each 
of them should be considered as indicative. 

A more detailed bi--dimensional fit of the galaxy images is under way, to
disentangle the different morphological component and to better
constrain the different components and therefore 
the SMBH mass (Donnarumma et al.
in preparation; Mignoli et al. in preparation).

\acknowledgements 
The original matter presented in this paper is the
result of the effort of a large number of people, in particular of the
{\tt HELLAS2XMM} team, which we warmely thank.  This work was partially
supported by the Italian Space Agency (ASI) under grant I/R/057/02, by
INAF under grant \# 270/2003 and MIUR grant Cofin--03--02--23.

\end{article}

\begin{thebibliography}{0}


\bibitem{ale01} 
Alexander D.M. \etal, \aj\ {\bf 122}, 2156 (2001). 

\bibitem{ale03} 
Alexander D.M. \etal, \aj\ {\bf 126}, 539 (2003). 

\bibitem{bal02} 
Baldi A., Molendi S., Comastri A., Fiore F., Matt G. Vignali C., 
\apj\ {\bf 564}, 190 (2002).

\bibitem{bru03} 
Brusa M., Comastri A., Mignoli M.,  \etal, \aap\ {\bf 409}, 65 (2003). 

\bibitem{}
Cimatti A., \etal, \aap\  {\bf 381}, L68 (2002).

\bibitem{com95}
Comastri A., Setti G. Zamorani G., Hasinger G., \aap\ {\bf 296}, 1 (1995). 

\bibitem{com95}
Comastri A., in "Supermassive Black Holes in the
Distant Universe", Ed. A. J. Barger. Kluwer Academic Publishers, 
Chapter 8, in press,  astro--ph/0403693 (2004).

\bibitem{}  
Elvis M. \etal, \apjs\ {\bf 95}, 1 (1994). 

\bibitem{}  
Fabian, A.C., in "Carnegie Observatories Astrophysics Series, Vol. 1: 
Coevolution of Black Holes and Galaxies," ed. L. C. Ho 
(Cambridge: Cambridge Univ. Press) astro--ph/0304122 (2003).

\bibitem{fio03} 
Fiore, F., Brusa M., Cocchia F., \etal, \aap\ {\bf 409}, 79 (2003). 

\bibitem{gia02} 
Giacconi, R.  \etal, \apjs\ {\bf 139}, 369 (2002). 

\bibitem{gia02} 
Hasinger G., in 
Proceedings of the conference: "The restless high energy universe",
held in Amsterdam, May 2003. To be published in: Nucl. Physics B. Suppl. Ser., E.P.J. van den Heuvel, J.J.M. in 't Zand, and R.A.M.J. Wijers (eds.).
astro--ph/0310804 (2003).

\bibitem{mig04} 
Marconi A., Hunt L.K., \apj\ {\bf 589}, L21 (2003).

\bibitem{mig04} 
Mignoli, M., Pozzetti L., Comastri A., \etal, \aap, in press, 
astro-ph/0401298 (2004). 

\bibitem{per04}
Perola, G.C., Puccetti S., Fiore F. \etal, \aap, in press, astro--ph/0404044 
(2004).  

\bibitem{per04}
Ueda Y., Akiyama M., Ohta K., Miyaji T. \apj\ {\bf 598}, 886 (2003).

\end{thebibliography}
\end{document}